\documentclass[a4paper,fleqn,usenatbib]{mnras}
\usepackage{mathptmx}

\usepackage[T1]{fontenc}
\usepackage{ae,aecompl}

\usepackage{url,hyperref}

\usepackage{graphicx}	
\usepackage{amsmath}	
\usepackage{amssymb}	

\title[VPOS and SDSS satellites]{The alignment of SDSS satellites with the VPOS: effects of the survey footprint shape}

\author[Marcel S. Pawlowski]{
Marcel S. Pawlowski,$^{1}$\thanks{E-mail: marcel.pawlowski@case.edu}\\
$^{1}$Department of Astronomy, Case Western Reserve University,
              10900 Euclid Avenue, Cleveland, OH, 44106, USA
}

\date{Accepted 2015 November 11.  Received 2015 November 10; in original form 2015 September 14}

\pubyear{2015}

\begin{document}
\label{firstpage}
\pagerange{\pageref{firstpage}--\pageref{lastpage}}
\maketitle

\begin{abstract}
It is sometimes argued that the uneven sky coverage of the Sloan Digital Sky Survey (SDSS) biases the distribution of satellite galaxies discovered by it to align with the polar plane defined by the 11 brighter, classical Milky Way (MW) satellites. This might prevent the SDSS satellites from adding significance to the MW's Vast Polar Structure (VPOS). We investigate whether this argument is valid by comparing the observed situation with model satellite distributions confined to the exact SDSS footprint area. We find that the SDSS satellites indeed add to the significance of the VPOS and that the survey footprint rather biases away from a close alignment between the plane fitted to the SDSS satellites and the plane fitted to the 11 classical satellites. Finding the observed satellite phase-space alignments of both the classical and SDSS satellites is a $\sim5\sigma$ event with respect to an isotropic distribution. This constitutes a robust discovery of the VPOS and makes it more significant than the Great Plane of Andromeda (GPoA). Motivated by the GPoA, which consists of only about half of M31's satellites, we also estimate which fraction of the MW satellites is consistent with being part of an isotropic distribution. Depending on the underlying satellite plane width, only 2 to 6 out of the 27 considered MW satellites are expected to be drawn from isotropy, and an isotropic component of $\gtrsim 50\,\%$\ of the MW satellite population is excluded at 95\,\% confidence.
\end{abstract}


\section{Introduction}
\label{sect:intro}

In answering a question posed by the French royal astronomical society, \citet{Bernoulli1734} was one of the first to apply probability theory to astronomy. He set out to determine whether the alignment of the then-known six planets and their orbits in the solar system along a common ecliptic plane could arise by chance, assuming they are drawn from isotropic distributions. If the observed arrangement were not unlikely to arise by chance, no particular formation mechanism for the planetary alignment would be required. Using several estimates he found the probability to be very low, between $2.7 \times 10^{-6}$\ and $7 \times 10^{-7}$. According to Bernoulli, ''[...] cette probabili\'e est si petite, quelle doit passer pour une impossibilit\'e morale''\footnote{''[...] this probability is so small that it must be received as a moral impossibility.''}. Consequently, a formation mechanism for coherently orbiting planetary systems had to be invoked. While the details of his estimations can be criticized, it is tempting to adopt his standard (and phrasing) of statistical significance.

Today, we face a similar challenge on a (spatially) much larger scale. The Milky Way (MW) is surrounded by a Vast Polar Structure (VPOS) of satellite objects \citep{Pawlowski2012a}. It has already been known for a long time that most of the brightest (''classical'') MW satellite galaxies are spatially distributed in a highly flattened configuration that is aligned with the Magellanic Stream \citep{KunkelDemers1976,LyndenBell1976}. Maybe more importantly, proper motion measurements for these objects have revealed that most of these satellites co-orbit in this planar structure \citep{Metz2009,PawlowskiKroupa2013}. Consequently, a certain resemblance to the planets in the solar system can not be denied, even though the alignment is less narrow and less coherent.

The origin of this satellite galaxy alignment has since been debated in the literature. \citet{Kroupa2005} were the first to argue that planar satellite distributions are in conflict with the expected distribution of primordial dwarf galaxies in the dark energy plus cold dark matter framework of cosmology ($\Lambda$CDM). This argumentation has since been both challenged \citep[e.g.][]{Zentner2005,Libeskind2009,Wang2012} and supported \citep{Metz2008,Ibata2014a,Pawlowski2014,PawlowskiMcGaugh2014b}. One of the earliest suggested (but non-cosmological) origins is that satellites distributed along a common plane around the MW originate from the breakup of a progenitor object such as the Magellanic Cloud \citep{LyndenBell1983}. This idea has prevailed in the debate and evolved into to the concept of Tidal Dwarf Galaxies \citep{Metz2007,Pawlowski2012a}. TDGs are second-generation dwarf galaxies formed from the debris of colliding disk galaxies, which are naturally phase-space correlated if they form out of a common tidal tail \citep{Wetzstein2007,Duc2011}. As a possible origin of planar satellite distributions they are actively discussed in an ever-increasing variety of dynamical and cosmological frameworks, including TDGs expelled from M31 within a model based on standard $\Lambda$CDM cosmology \citep{Fouquet2012,Hammer2013,Yang2014}, TDGs from an encounter between the MW and M31 \citep{Zhao2013} as expected in Modified Newtonian Dynamics \citep[MOND][]{Milgrom1983}, and TDGs in models of non-standard dark matter \citep{Foot2013,RandallScholtz2014}.

In this ongoing debate on the origin of the VPOS and its implications for cosmology, most contributions limit themselves to the empirical information provided by only the 11 classical satellites, even though the discovery of additional, fainter MW satellites can provide further clues and support of the VPOS phenomenon if they are preferentially aligned with the plane defined by the 11 classical satellites.

Indeed, the fainter satellite galaxies discovered since 2005 (see Table \ref{tab:satellites} for references) were not only found to lie close to the plane defined by the classical satellites \citep{Metz2009,Pawlowski2015a}, they even independently define a very similar plane \citep{Kroupa2010}. However, it is frequently cautioned that \citep[until recently, see][]{DES2015,Koposov2015,KimJerjen2015a,Martin2015,DES2015b} most of the faint and ultra-faint satellites of the MW were discovered in the Sloan Digital Sky Survey \citep[SDSS;][]{York2000}, which does not have an even sky coverage but was initially focussed on a region $\sim 60^{\circ}$\ around the northern Galactic pole. It is thus plausible to argue that this biases the discovered satellites to align with a \textit{polar} structure, which in turn might reduce or even nullify the information the fainter satellites provide on the VPOS phenomenon. This qualitative argument, however, needs to be tested quantitatively.

We therefore set out to test whether the shape and orientation of the SDSS footprint dominates the distribution of satellites discovered in this survey by biasing them to align with the plane of classical satellites to a degree which prohibits them from providing any conclusive evidence in regard to the VPOS. By constructing model satellite distributions constrained to the SDSS footprint, we ensure the most detailed modelling of the influence of the SDSS survey area on the flattening of the MW satellite distribution to date.

In recent years, a growing number of satellite alignments similar to the VPOS have been reported (see \citealt{Ibata2014b} and \citealt{Tully2015}, and references in \citealt{PawlowskiKroupa2014}). The most prominent and significant discovery is certainly the Great Plane of Andromeda (GPoA). \citet{Ibata2013} found that 15 out of 27 satellites of M31 can be assigned to a very narrow plane seen almost edge-on from the MW. The line-of-sight velocities of these satellites reveal that 13 out of the 15 might have the same orbital sense. This ''Great Plane of Andromeda'' (GPoA) is thus similar to the VPOS, but consists of only a subset of all known M31 satellites, which motivates us to estimate what fraction of the known MW satellite galaxies might be part of an isotropic population in addition to a planar one.

Our paper is structured as follows. Sect. \ref{sect:observed} summarises the sample of observed MW satellites which will be considered in this work, Sect. \ref{sect:model} presents the algorithm generating model satellite realisations from underlying isotropic and planar distributions. The effect of the SDSS footprint and the significance of the VPOS including the SDSS satellites is determined in Sect. \ref{sect:significance}, while the fraction of isotropically distributed satellites consistent with the observed situation is estimated in Sect. \ref{sect:isofraction}. In Sect. \ref{sect:conclusion} our results are summarized and discussed.


\section{Observed MW satellites}
\label{sect:observed}

\begin{table}
	\centering
	\caption{Classical and SDSS satellites around the MW, their Galactocentric distance $r_{\mathrm{MW}}$\ in kpc, SDSS data release in which this object was discovered ($^{*}$ marks discoveries made in the SEGUE imaging footprint) and references to the first publication mentioning each object.
	}
	\label{tab:satellites}
	\begin{tabular}{lccl} 
		\hline
        Name & $r_{\mathrm{MW}}$ & SDSS & Reference\\
		\hline
\textit{Classical} & & &\\
Sagittarius &   18 &  --- & \citealt{Ibata1994} \\
LMC &   50 &  --- & --- \\ 
SMC &   61 &  --- & --- \\ 
Draco &   76 &  --- & \citealt{Wilson1955} \\
Ursa Minor &   78 &  --- & \citealt{Wilson1955} \\
Sculptor &   86 &  --- & \citealt{Shapley1938a} \\
Sextans &   89 &  --- & \citealt{Irwin1990} \\
Carina &  107 &  --- & \citealt{Cannon1977} \\
Fornax &  149 &  --- & \citealt{Shapley1938b} \\
Leo II &  236 & --- & \citealt{HarringtonWilson1950} \\
Leo I &  257  &  --- & \citealt{HarringtonWilson1950} \\ \hline
\textit{SDSS} & & &\\
Segue I &   28 & DR6$^{*}$ & \citealt{Belokurov2007} \\
Ursa Major II &   38 &  DR4 & \citealt{Zucker2006b} \\ 
Bootes II &   39 &  DR5 & \citealt{Walsh2007} \\
Segue II &   41 & DR7$^{*}$ & \citealt{Belokurov2009} \\ 
Willman 1 &   43 &  DR2 & \citealt{Willman2005a} \\ 
Coma Berenices &   45 & DR5 & \citealt{Belokurov2007} \\ 
Bootes III &   46 &  DR5 & \citealt{Grillmair2009} \\ 
Bootes &   64 &  DR5 & \citealt{Belokurov2006} \\ 
Ursa Major &  102 &  DR2 & \citealt{Willman2005b} \\ 
Hercules &  126 &  DR5 & \citealt{Belokurov2007} \\ 
Leo IV &  155 &  DR5 & \citealt{Belokurov2007} \\ 
Canes Venatici II &  161 &  DR4 & \citealt{SakamotoHasegawa2006} \\ 
                   &      &  DR5 & \citealt{Belokurov2007} \\
Leo V &  179 &  DR6 & \citealt{Belokurov2008} \\ 
Pisces II &  181 &  DR7$^{*}$ & \citealt{Belokurov2010} \\ 
Pegasus III &  203 &  DR10 & \citealt{KimJerjen2015} \\
Canes Venatici &  218 &  DR5 & \citealt{Zucker2006a}\\ \hline
	\end{tabular}
\smallskip
\end{table}

The 11 brightest satellites (upper part of Table \ref{tab:satellites}) are commonly referred to as the classical satellites. They all are known for more than 20 years, and it is generally assumed that no object of similar luminosity remains undiscovered around the MW (except if hidden right behind the MW disk due to dust obscuration or stellar crowding). The 11 classical satellites have a highly flattened spatial distribution \citep{Kroupa2005}, characterized by the root-mean-square (rms) height from a plane fitted to their positions ($\Delta_{\mathrm{rms}}^{\mathrm{class}} = 19.6\,\mathrm{kpc}$), or the short-to-long axis ratio ($(c/a)^{\mathrm{class}} = 0.18$). Within their uncertainties, the orbital poles (directions of angular momenta) of at least eight of the classical satellites are strongly concentrated \citep{PawlowskiKroupa2013}, which can be characterized by the spherical standard deviation from their average direction ($\Delta_{\mathrm{std}} = 27.2^{\circ}$\ using the updated proper motion of \citet{Pryor2015}).

In the data collected by the SDSS, 16 fainter MW satellite galaxies have been discovered by searches for stellar over-densities. The majority of these were found between 2005 and 2010 in Data Releases (DR) 4 to 7 \citep{DR4,DR5,DR6,DR7}, but recently one additional faint and distant MW satellite, Pegasus\,III, was discovered by \citet{KimJerjen2015} using DR10 \citep{DR10}. A plane fitted to these 16 SDSS satellite has an rms height of $\Delta_{\mathrm{rms}}^{\mathrm{SDSS}} = 25.9\,\mathrm{kpc}$, a short-to-long axis ratio of $(c/a)^{\mathrm{SDSS}} = 0.26$ and is inclined by only $\theta_{\mathrm{class}}^{\mathrm{SDSS}} = 22^{\circ}$\ from the plane fitted to the 11 classical satellites. Taken together, the full sample of 27 MW satellites has a rms height of $\Delta_{\mathrm{rms}}^{\mathrm{all}} = 28.0\,\mathrm{kpc}$ and an axis ratio of $(c/a)^{\mathrm{all}} = 0.27$.

In contrast to previous analyses \citep[e.g.][]{Pawlowski2013}, Ursa Major I is not considered in the current analysis. The dominating reason is that it was not discovered in the SDSS survey but in the 2 Micron All Sky Survey (2MASS) by \citet{Martin2004}. Another argument against inclusion in our analysis is that the nature of Ursa Major I is uncertain, with some authors arguing that it is a a satellite galaxy \citep{Martin2004}, and others preferring an interpretation as a MW disk substructure \citep{Momany2006}. Since the object is situated very close to the MW it does not substantially affect satellite plane fits.

In contrast to Ursa Major I, the Pisces Overdensity (or Pisces I if interpreted as a disrupted dwarf galaxy) was discovered using data of the SDSS survey \citep{Watkins2009}. However, we decided to not consider it in this analysis because it was discovered using a different technique (the clustering of RR Lyra stars identified in SDSS Stripe 82), which is not applicable to the other directions covered by SDSS since the required repeated observations are not available. We note, however, that the Pisces Overdensity lies well within the VPOS and can therefore be expected to add to this structures' significance.

In the present study, we do not consider the more recently discovered faint MW satellite candidates from a variety of surveys. A major reason for this is that many discovered objects still lack follow-up observations clarifying whether they are satellite galaxies or star clusters (a distinction which in becomes increasingly blurry, too). Furthermore, these surveys are not yet completed, and some have much smaller sky coverage than SDSS and/or are pointing close to the Magellanic Clouds (MCs). The latter makes it difficult to discriminate between MW and potential current or former MC satellites, constituting an additional bias to studies of satellite alignments because the MCs are part of and orbit within the VPOS

\section{Models}
\label{sect:model}

\begin{figure}
   \centering
   \includegraphics[width=85mm]{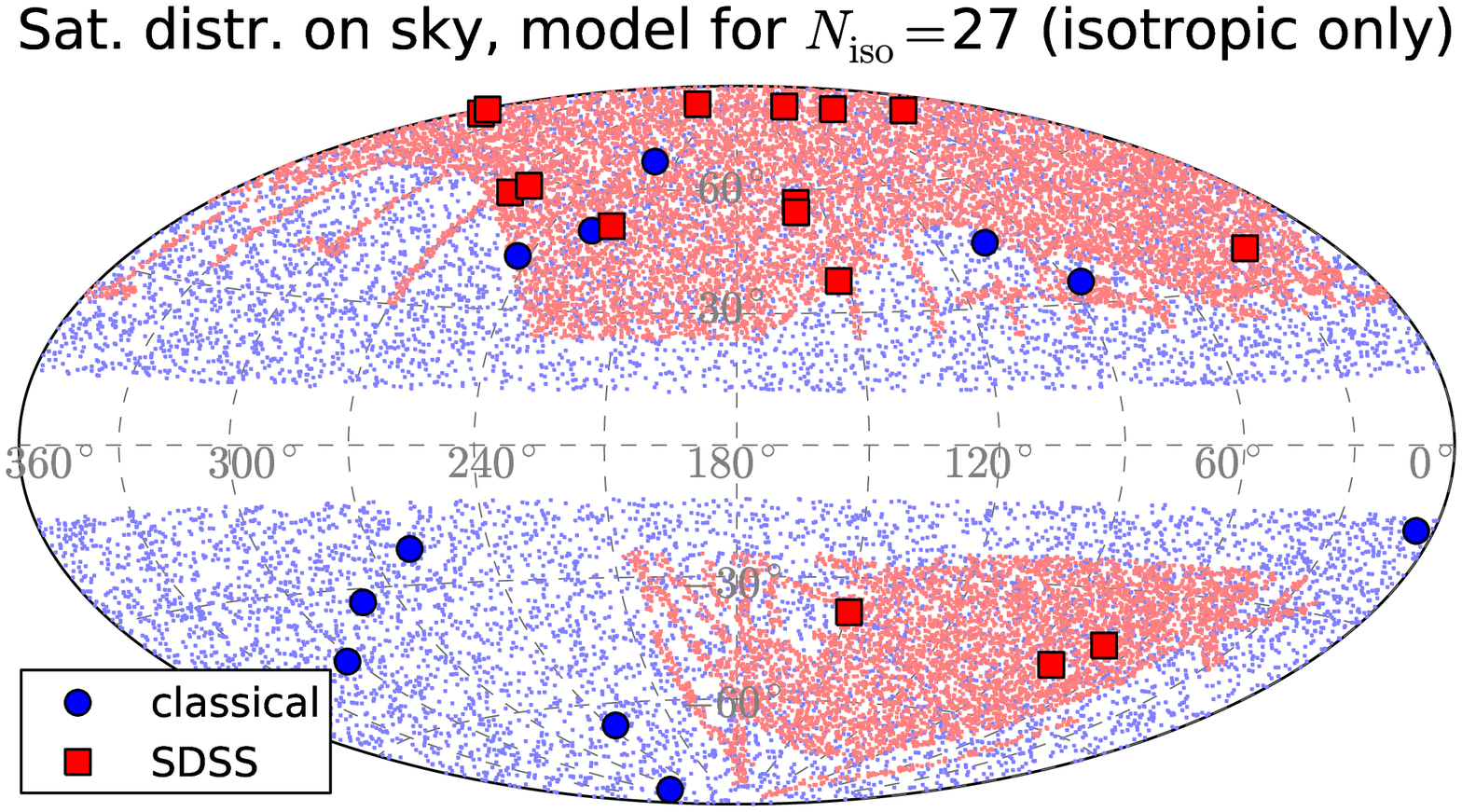}
   \includegraphics[width=85mm]{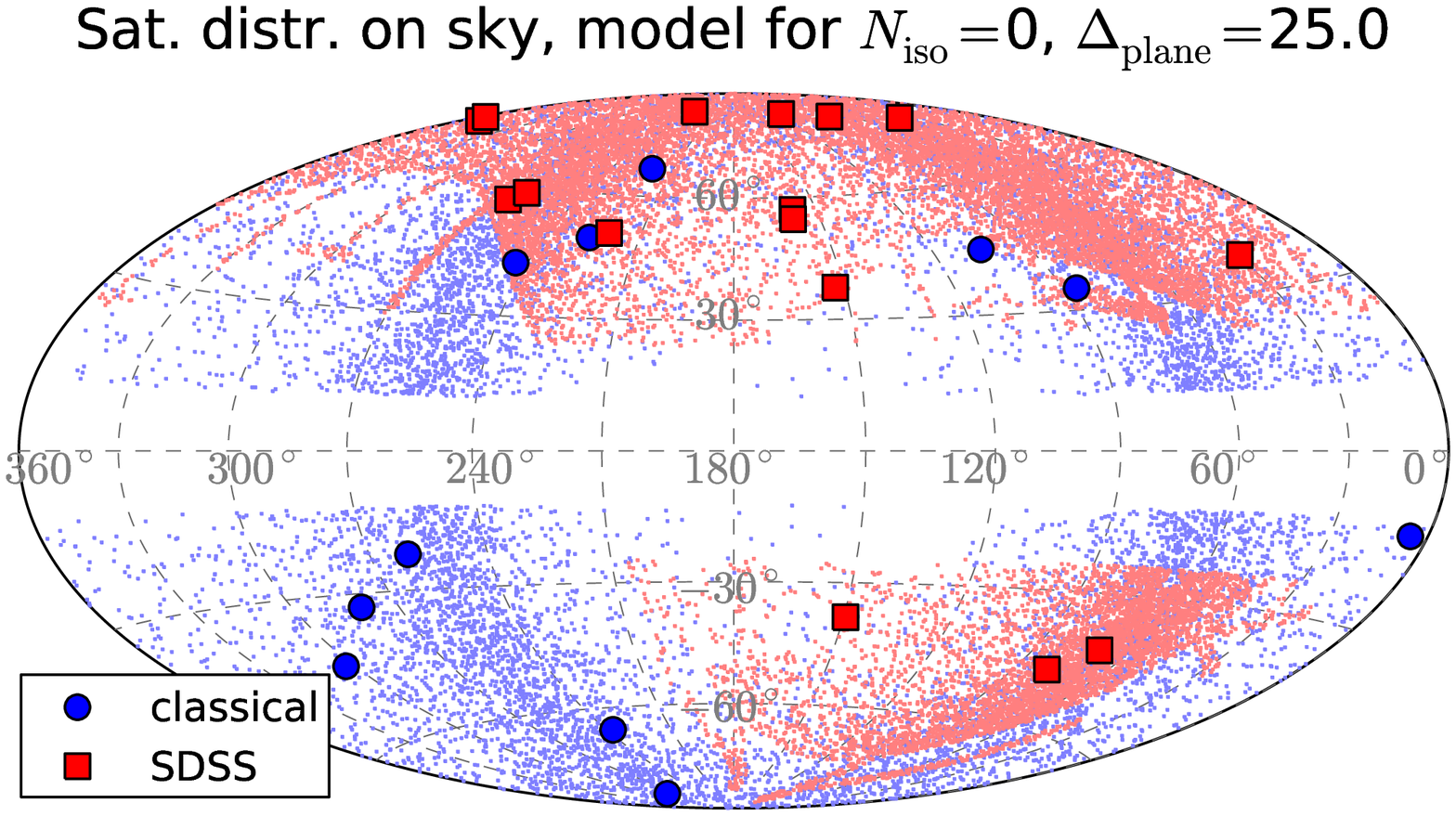}
   \caption{
Distribution of observed MW satellites (symbols) and model satellites (points) for 1000 realisations each in Galactic coordinates. The \textit{top panel} is for an isotropic model distribution, the \textit{bottom panel} shows a purely planar model distribution. The plots reveal the different assumed obscuration regions for the classical (blue, $|b| \leq 12^{\circ}$) and faint (red, $|b| \leq 24^{\circ}$) model distributions, and show that the faint satellites are confined to the area that was covered by SDSS DR10.
   }
              \label{fig:ASP}
\end{figure}

\subsection{Isotropic satellite realisations}
\label{sect:modeliso}

To determine the significance of the flattening of the MW satellite system, we compare the observed characteristics of the VPOS with satellite distributions drawn from isotropy\footnote{This is different from the probability to find such a distribution among sub-halo based satellite systems in $\Lambda$CDM models, which have a slightly more anisotropic distribution \citep{Wang2013, PawlowskiMcGaugh2014b}. We will present such a comparison to cosmological simulation in a future contribution.}. 

We construct a total of 50000 realisations, each consisting of equivalents of the 11 classical and 16 SDSS satellites. For each satellite, we draw a random angular position on a unit sphere around on the Galactic center, but retain its Galactocentric distance $r_{\mathrm{MW}}$. The position vector is then transformed to a Heliocentric coordinate system (assuming the Galactocentric distance of the Sun to be 8.3\,kpc, \citealt{McMillan2011}). 

For the equivalents of the classical satellites, we assume that the obscuration by the Galactic disk hinders any detection below a Galactic latitude of $|b| \leq 12^{\circ}$. We fix this obscuration angle empirically by requiring that our isotropic models are likely to contain a satellite at a comparably low latitude as Sagittarius (at $b = -14.2^{\circ}$, Carina has the next-smallest $b$ of $-22.2^{\circ}$). For an isotropic distribution of 11 satellites we expect that there is on average one satellite between $|b| = 12^{\circ}$\ and $|b| = \arcsin{\left(1 / 11 + \sin{12^{\circ}}\right)} = 17.4^{\circ}$, thus our choice is consistent with finding an object like Sagittarius.

If the randomly chosen position lies within this obscured region, a new random position is generated until one is found that is not obscured. Once positions for each of the 11 classical satellites are found, we also construct their orbital poles by drawing random velocity direction from an isotropic distribution.

For the SDSS satellites, the procedure differs in two aspects. Due to their lower surface brightness the angle defining the obscured region is increased by a generous (and thus conservative) factor of two, thus no SDSS model satellites within $|b| \leq 24^{\circ}$\ are allowed. This cut in galactic latitude agrees with that used by \citet{Hargis2014}, and like them we assume that the average detectability of the satellites in SDSS is not affected by their galactic latitude, which is in line with the findings of \citet{Walsh2009}. More importantly, the randomly drawn (Heliocentric) positions must fall within the SDSS survey region. Since all 16 satellites have been discovered in SDSS DR10 or earlier (see Table \ref{tab:satellites}) we consider the area covered up to that DR in our analysis\footnote{Performing the analysis using only the 15 satellites discovered within DR7 or earlier data and restricting the SDSS region to the DR7 footprint provides similar results.}. In order to be as close as possible to the original survey coverage, we compare whether a position lies within one of the $2.5^{\circ}$\ wide SDSS runs. This assumes that a satellite can be discovered if $\geq 50$\,per cent of its surface fall within the imaging footprint of SDSS. The DR10 run parameters are extracted from the photoRunAll-dr10.par file available from the SDSS website\footnote{https://www.sdss3.org}. While not very efficient, this procedure maximizes the accuracy of the footprint modelling. If a randomly chosen position lies outside of any SDSS stripe, a new position is generated and the procedure is repeated until a position is found which fulfils this requirement. As a control-sample to test what effects the survey shape has on the satellite distribution we also generate 50000 realisations in which the SDSS footprint is ignored, i.e. the SDSS satellites are drawn from an isotropic distribution but confined to $|b| \geq 24^{\circ}$.

The upper panel of Fig. \ref{fig:ASP} illustrates the resulting distribution for 1000 realisations of the 11 classical (blue) and 16 SDSS satellites (red). The plot shows that some of the SDSS runs extend beyond the regions of continuous coverage. For each realisation planes are fitted to satellite positions of the three samples (the equivalents of the 11 classical and 16 SDSS satellites, and to the total sample of 27 satellites) using the technique summarized in \citet{Pawlowski2013}. The plane fit parameters (rms height $\Delta_{\mathrm{rms}}$, minor-to-major axis ratio $c/a$, inclination between the classical and the modelled SDSS satellite plane $\theta_{\mathrm{class}}^{\mathrm{SDSS}}$) are recorded. 
For the classical satellites, also the spherical standard deviation $\Delta_{\mathrm{sph}}$\ of the eight most-concentrated orbital poles (i.e. calculated for that combination of 8 out of 11 model satellites resulting in the smallest $\Delta_{\mathrm{std}}$) is calculated (using the technique outlined in \citealt{PawlowskiKroupa2013}) and recorded.

\subsection{Combinations of isotropic and planar realisations}
\label{sect:modelplanes}

\begin{figure*}
   \centering
   \includegraphics[width=58mm]{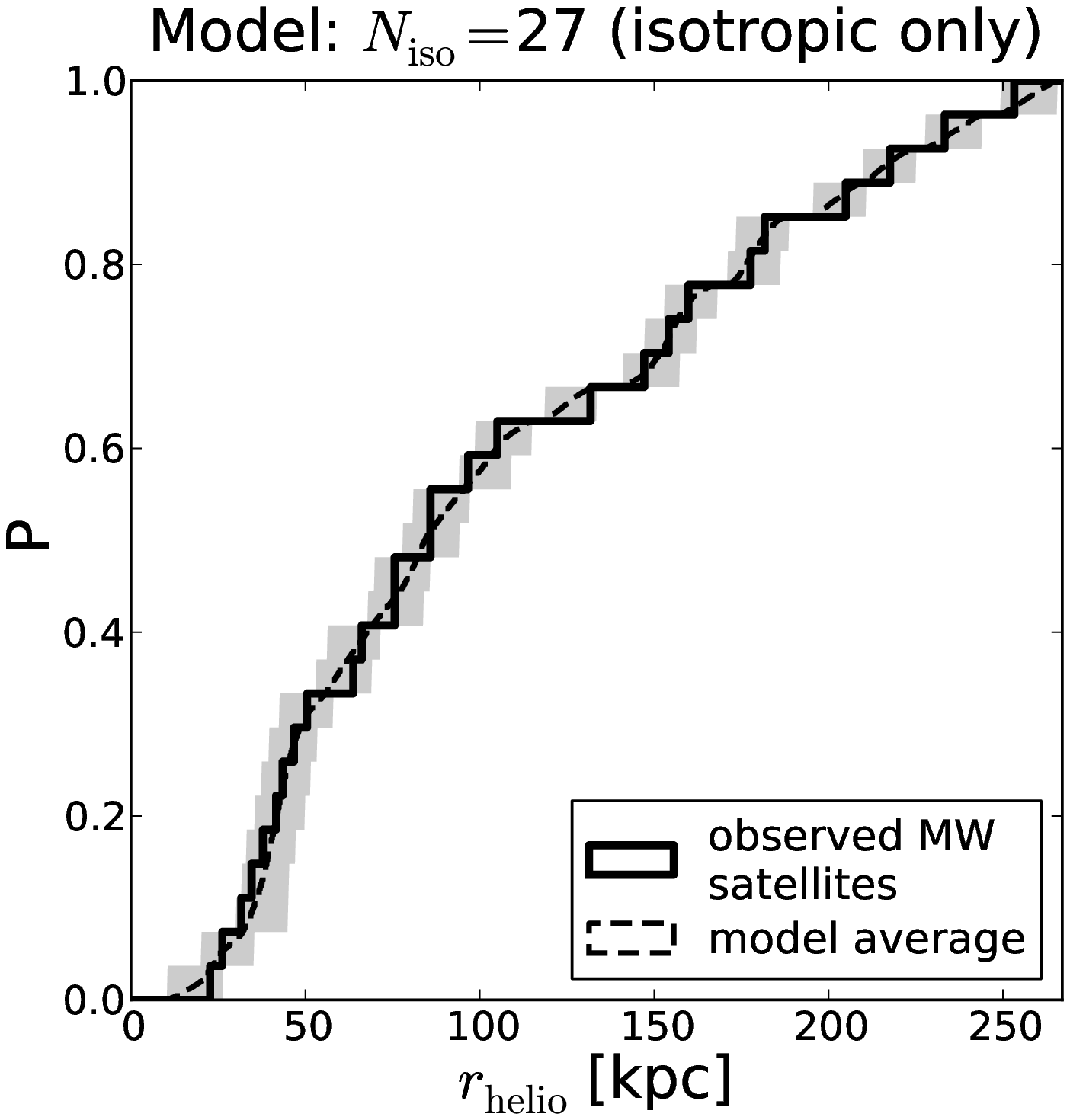}
   \includegraphics[width=58mm]{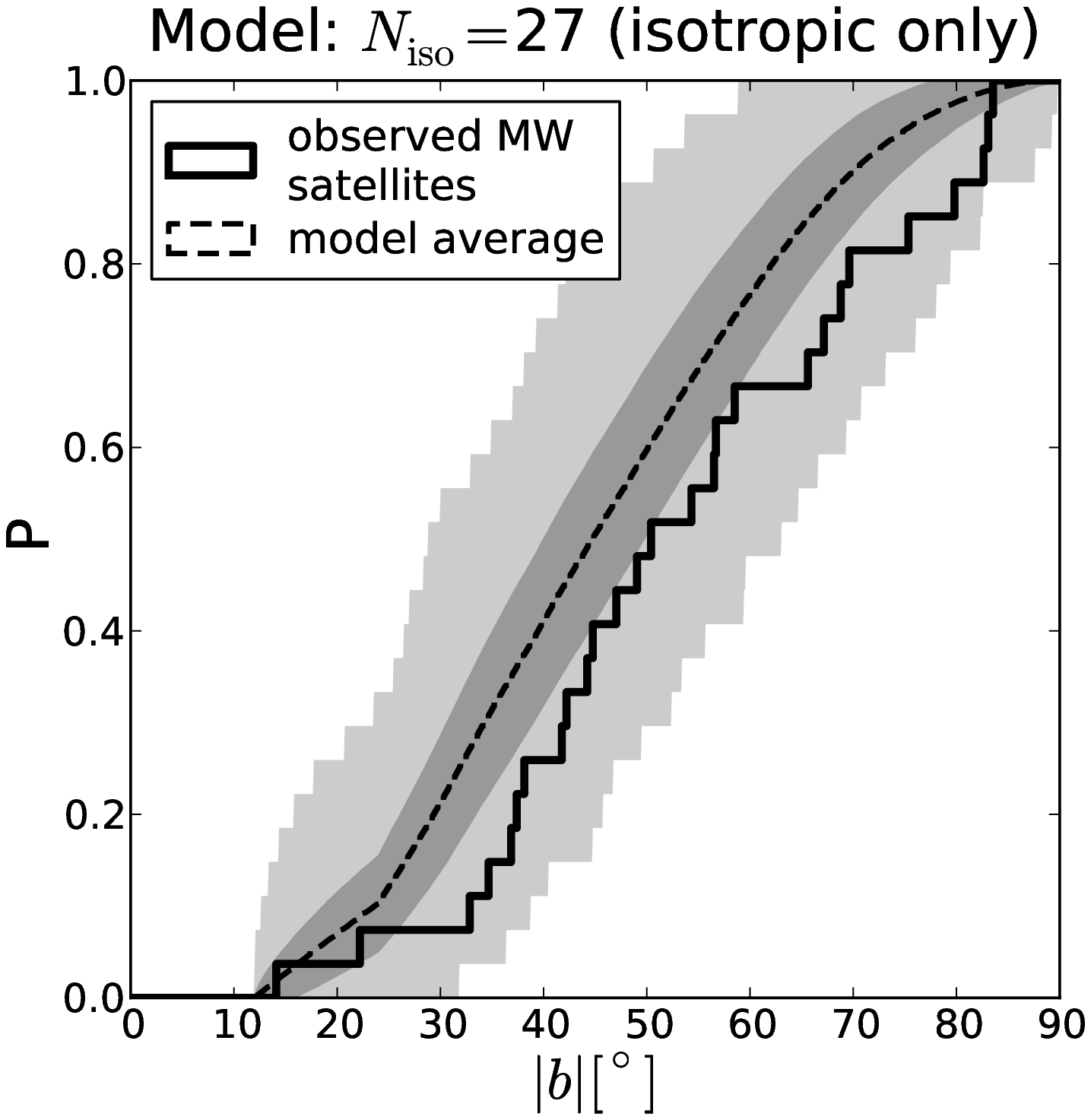}
   \includegraphics[width=58mm]{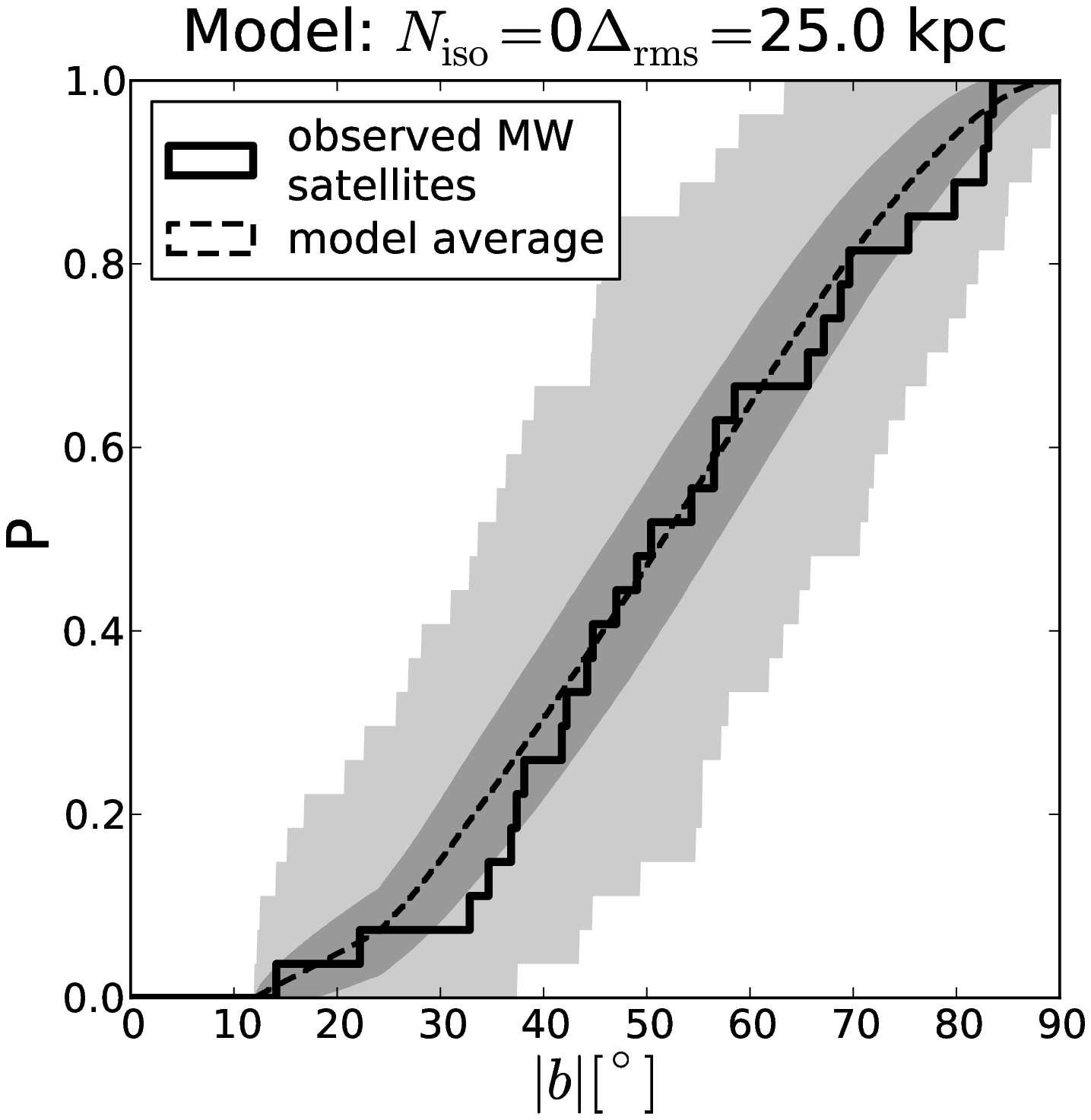}
   \caption{
\textit{Left panel}: Cumulative distribution of Heliocentric distances $r_{\mathrm{helio}}$\ for the observed MW satellites (solid line) and the average for 5000 isotropic realisations. The grey shaded area indicates the maximum variation among the 5000 models. Our model realisations reproduce the observed distribution very well. The other two panels compare the cumulative distribution of absolute Galactic latitude $|b|$\ for the observed MW satellites (solid line) with the average for the 5000 isotropic realisations (\textit{middle}) and with 5000 realisations in which all satellites are part of a polar plane with rms height of $\Delta_{\mathrm{rms}}^{\mathrm{input}} = 25$\,kpc (\textit{right}). The dark grey shaded region indicates the $1\sigma$\ scatter around this mean and the light grey shaded region indicates the range between the maximum and minimum. 
   }
              \label{fig:distlat}
\end{figure*}

To estimate which fraction of the known MW satellites might be part of an isotropic distribution, we also construct model satellite distributions in which only some of the satellites are drawn from isotropy, while the others are part of an artificial planar distribution. The input plane passes through the Galactic center and its normal direction points to $(l, b) = (160^{\circ}, 0^{\circ})$, an orientation similar to that of the observed VPOS\footnote{We have checked that varying the plane orientation within the range allowed by the uncertainties on the observed VPOS does not affect our results.}.

The number of isotropically distributed model satellites is $N_{\mathrm{iso}}$. For each realisation, a random sub-sample of $N_{\mathrm{iso}}$\ out of the full sample of all classical and SDSS satellites are drawn and their positions are randomized as described in Sect. \ref{sect:modeliso}. This procedure thus assumes that the classical and the fainter satellites have the same chance to be part of the isotropic distribution, which is consistent with the finding that the M31 satellites in and out of the GPoA share the same mix of properties \citep{Collins2015}.

The positions of the remaining $27 - N_{\mathrm{iso}}$\ satellites that are modelled as part of a satellite plane are constructed as follows. Each satellite again retains its Galactocentric distance $r_{\mathrm{MW}}$. We draw the perpendicular offset of the satellite from the artificial plane using a Gaussian distribution with an input plane height $\Delta_{\mathrm{rms}}^{\mathrm{input}}$. If the resulting offset is larger than $3 \Delta_{\mathrm{rms}}^{\mathrm{input}}$, or if the Galactocentric distance of the satellite is smaller than the offset, then a new offset is drawn until it is below both of these limits. The offset from the plane and the Galactocentric distance of the corresponding satellite then define the angle of the satellite out of the plane. With this, two out of three coordinates -- the radial distance and one angular position -- for the modelled satellite are fixed. The remaining third coordinate, the azimuthal angle along the plane, is drawn from a flat distribution between 0 and $360^{\circ}$, thus assuming that the plane is evenly populated in all directions\footnote{Note that this is not the case for the GPoA plane, which is lopsided in the sense that 13 of 15 satellites are in the hemisphere pointing towards the MW \citep{Ibata2013,Conn2013}. We do not see clear indications for a similar lopsidedness in the MW satellite distribution, such that we assume the simpler model of an even distribution for this study.}. 

If the resulting position (in Heliocentric coordinates) is within the region obscured by the MW disk, or not within the SDSS survey area if the corresponding MW satellite was discovered in SDSS, then the whole procedure is repeated until an acceptable position is found.

For each $N_{\mathrm{iso}}$\ between 0 (purely planar) and 27 (purely isotropic) we construct 5000 realisations for input plane heights $\Delta_{\mathrm{rms}}^{\mathrm{input}}$\ between 5 and 50\,kpc, sampled in steps of 5\,kpc. In total, we thus construct $28 \times 10 = 280$\ sets of 5000 model realisations. As for the isotropic models, for each realisation planes are fitted to the satellite positions and their parameters are recorded. To illustrate the resulting satellite distribution on the sky for a purely planar case, the satellite positions of 1000 realisations with $\Delta_{\mathrm{rms}}^{\mathrm{input}} = 25$\,kpc are plotted in the lower panel of Fig. \ref{fig:ASP}.

\subsection{Consistency Checks}

While our procedure conserves the Galactocentric distance distribution of the MW satellites by definition, the Heliocentric distances do not stay at their exact observed values. Large differences in the Heliocentric distances between the observed and the model satellite positions would be problematic, because this could affect the detectability of the ultra-faint satellites. However, we have checked that this variation is minor, the cumulative distribution of model satellite distances closely follows the observed one (left panel of Fig. \ref{fig:distlat}).

The middle panel of Fig. \ref{fig:distlat} compares the cumulative distributions of absolute Galactic latitude $|b|$\ for satellites drawn from an isotropic distribution with the observed satellite positions. There is some discrepancy, the curve of observed latitudes mostly falls below the $1\,\sigma$\ region (i.e. the observed distribution is more concentrated towards the poles than the modelled one). This offset is most-likely caused by the difference in the underlying distribution, not inaccurate modelling of the sky coverage. Most of the observed MW satellites are distributed in a polar plane with a root-mean-square height of 20 to 30\,kpc, such that it is incorrect to compare with an underlying isotropic distribution. Instead, drawing the satellites from a planar distribution with a rms height comparable to the observed VPOS results in a good agreement of the cumulative latitude distribution with the $1\,\sigma$ region (right panel of Fig. \ref{fig:distlat}). This way the cumulative latitude distribution acts as another indication that the underlying MW satellite distribution is planar rather than isotropic, and confirms that our choice of Galactic disk obscuration and SDSS footprint provides realistic results.

A consequence of this is that estimates of the number of satellites that are obscured by the MW disk might be affected by incorrectly assuming isotropy for the underlying satellite distribution, resulting in an over-estimate of the number of obscured satellites. For example, \citet{Willman2004} have estimated the number of obscured MW satellites by pointing out that 9 out of the 11 classical satellites are at $|b| \geq 30^{\circ}$, which corresponds to only half the sky. Assuming isotropy, in which case the same number of satellites are to be expected within and outside of $30^{\circ}$ of the MW disk, the authors predict that $(9 / 50\,\%) - 11 = 7$\ satellites as luminous as the classical ones might be hidden behind the Galactic disk. However, our results illustrate that the expected latitude distribution depends on the assumed underlying satellite distribution. For a perfectly polar, narrow and evenly populated plane, the distribution in $b$\ would be uniform, so the region $|b| \geq 30^{\circ}$\ should contain two thirds of the satellites. In this case an estimate equivalent to that of \citet{Willman2004} predicts only $(9 / 67\,\%) - 11 = 2.5$\ satellites to be obscured. Our assumed obscuration region of $12^{\circ}$\ is consistent with this: $\frac{12}{90} = 13.3$\,per cent of a perfect polar plane are hidden. The expected total number of satellites then is $\frac{11}{(1.0 - 0.133)} = 12.7$, such that about 2 satellites are obscured in the assumed region.


\section{Significance of the VPOS}
\label{sect:significance}

\begin{figure*}
   \centering
   \includegraphics[width=58mm]{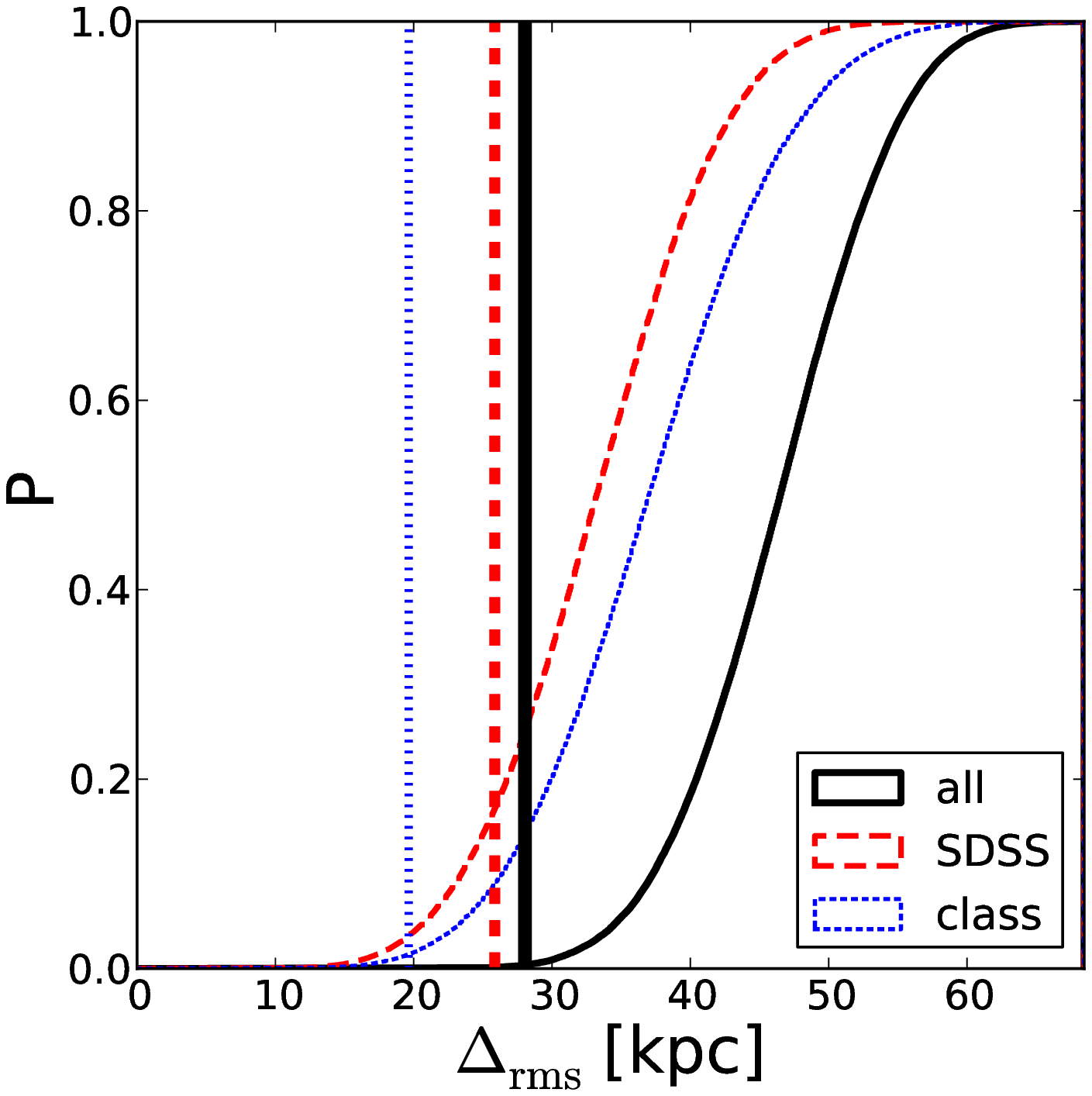}
   \includegraphics[width=58mm]{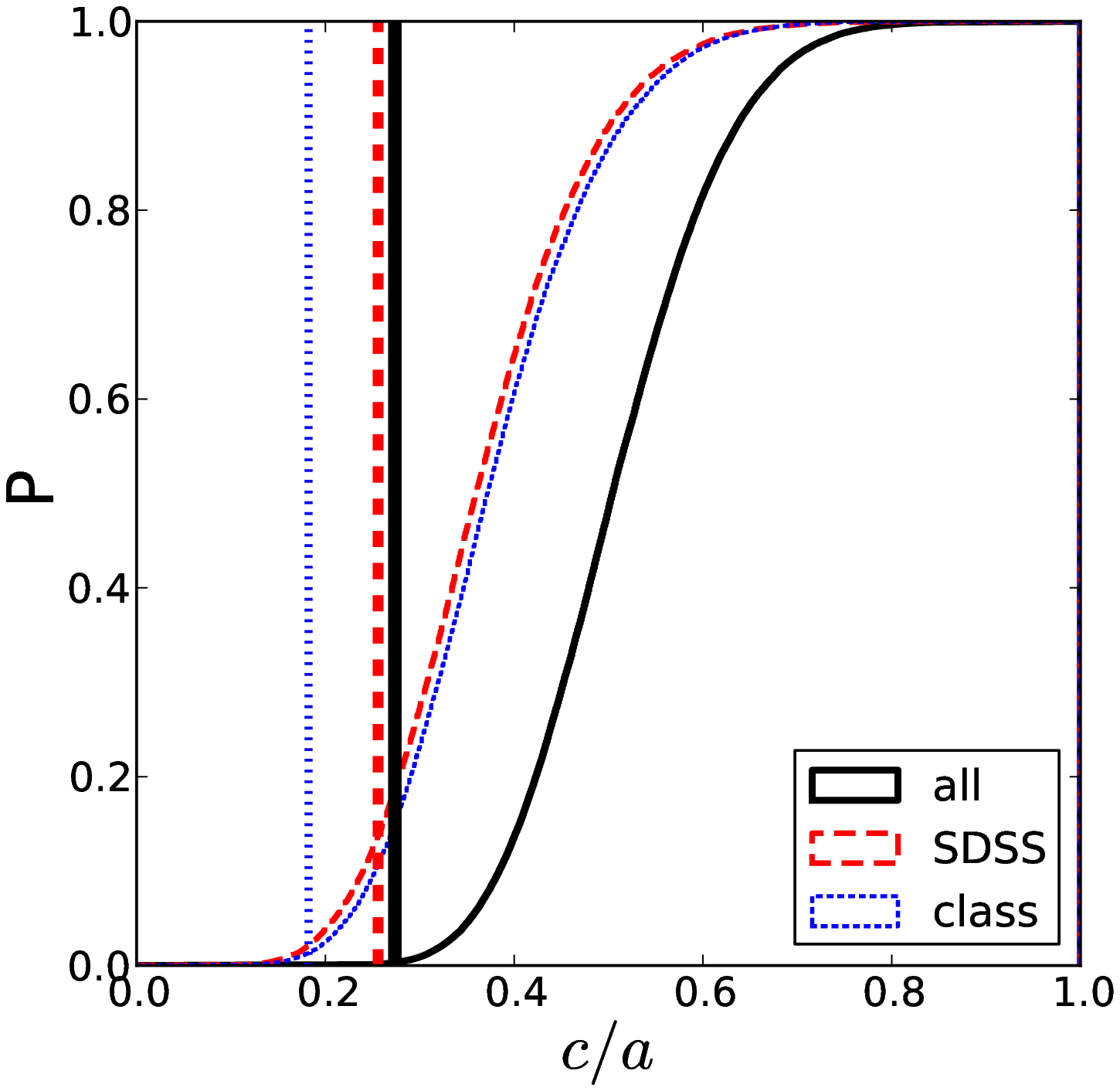}
   \includegraphics[width=58mm]{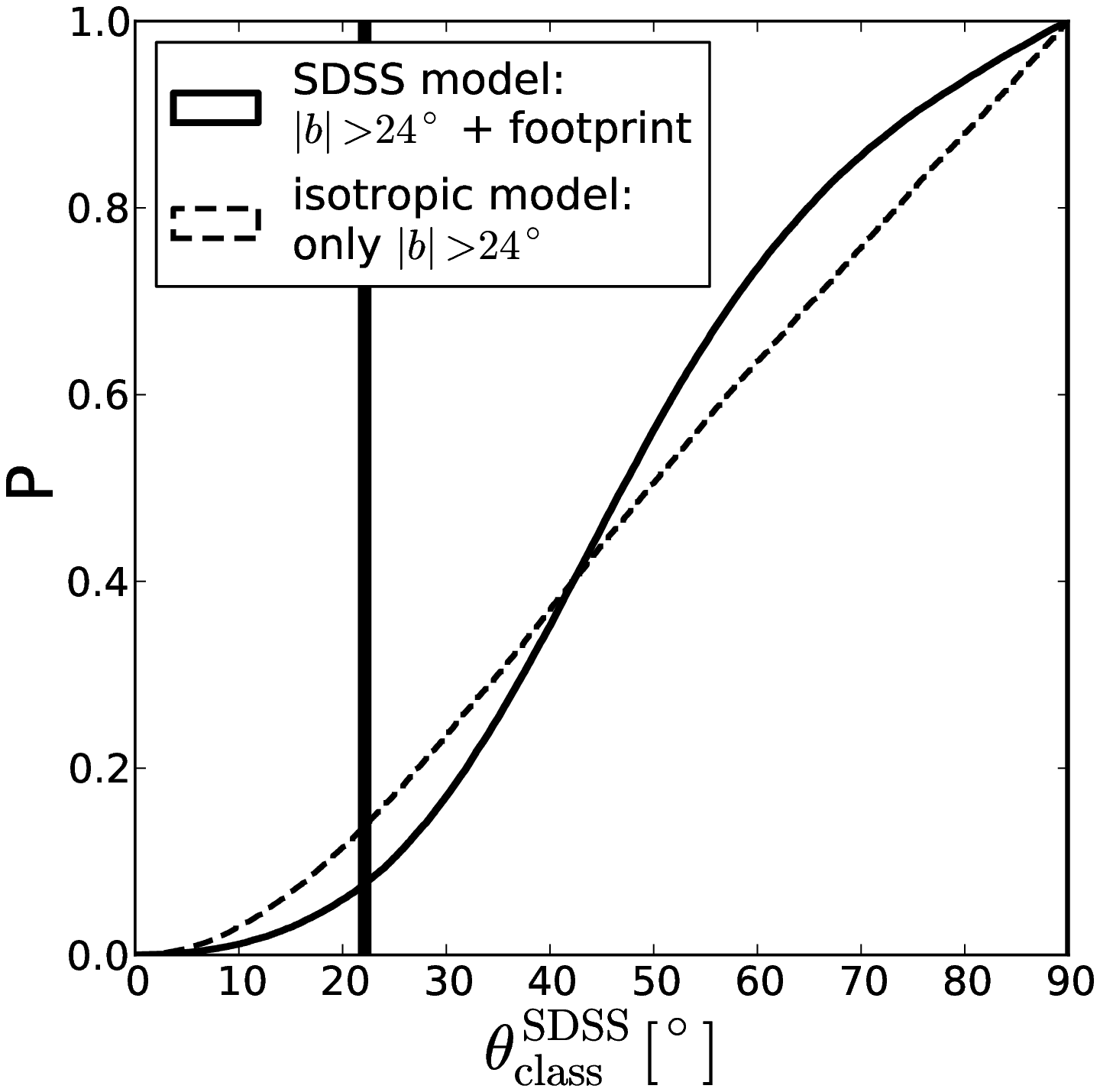}
   \caption{
Cumulative distributions, derived from 50000 isotropic realisations, of the rms heights of planes fitted to the satellite positions (\textit{left panel}), the axis ratios $c/a$\ (\textit{middle panel}) and the angle (\textit{right panel}) between the normal to the plane fitted to the observed classical MW satellites and the normal vectors to the planes fitted to the 16 SDSS satellite equivalents confined to the SDSS footprint (solid line) or ignoring the SDSS footprint (but not the Galactic latitude criterion, dashed line). The first two panels show the distributions for the 11 classical (dotted blue lines), 16 SDSS (dashed red lines) and all 27 satellites together (solid black lines). The values for these parameters derived from the observed MW satellites are indicated as vertical lines of the same colour and line style.
   }
              \label{fig:results}
\end{figure*}

\begin{figure*}
   \centering
   \includegraphics[width=150mm]{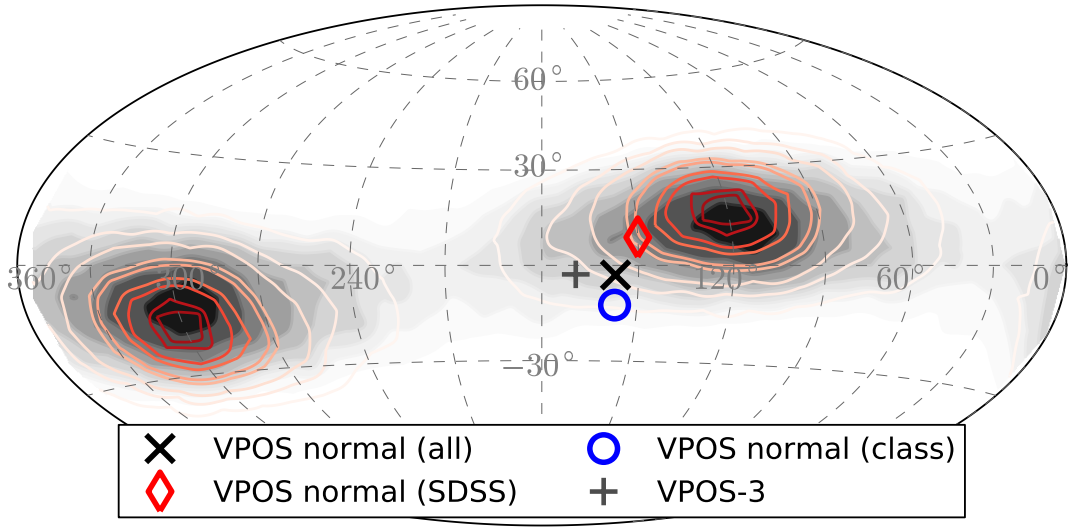}
   \caption{
Distribution of plane normal vectors for 50000 purely isotropic realisations constructed as described in Sect. \ref{sect:modeliso}. The grey filled contours illustrate the plane normal vector distribution for planes fitted to all 27 model satellites in each realisation, the red contour lines illustrate the normal vector distribution of planes fitted to the sub-sample of 16 SDSS satellite equivalents in each realisations. The regions enclosed by each subsequent contour line include 10\,per cent more plane normals.
The normal direction for the sub-sample of 11 classical MW satellite equivalents is not plotted, it is symmetric along Galactic latitude, with a preference for areas closer to the equator due to the assumed obscuration by the Galactic disk.
Also shown are the plane normal directions for the 11 classical satellites (blue circle), the 16 SDSS satellites (red diamond), both groups together (black cross), and the sample called VPOS-3 from \citet{Pawlowski2013}, which excludes three outliers (grey plus sign). The observed VPOS normals are about $30^{\circ}$\ to $50^{\circ}$\ away from the most-likely direction defined by the SDSS survey area.
   }
              \label{fig:ASPnormals}
\end{figure*}

We quantify the significance of a satellite structure by measuring the frequency $P$\ that satellite distributions drawn from isotropy are at least as extremely aligned as the observed MW satellites. This significance is either reported as a fraction or as the equivalent (two-sided) $\sigma$\ deviation for a normal distribution.

For the sake of completeness we perform the analysis using both the absolute plane rms heights $\Delta_{\mathrm{rms}}$\ ($P_{\mathrm{rms}}$) and the axis ratios $c/a$\ ($P_{c/a}$). This also acts as a consistency test, since both measures of the flattening of a satellite system should give comparable results, in particular because our analysis exactly preserves the observed radial distribution of the satellites from the MW. The resulting cumulative probabilities of plane heights are shown in the left (for $\Delta_{\mathrm{rms}}$) and middle (for $c/a$) panels of Fig. \ref{fig:results} for the fits to the model-equivalents of the 11 classical satellites (blue dotted lines), the 16 SDSS equivalents (red dashed lines) and the combined sample of 27 MW-satellite equivalents (black solid lines).

\subsection{Positions of the 11 classical satellites}

The frequency of the observed degree of flattening of 11 classical satellites alone is only $P_{\mathrm{rms}}^{\mathrm{class}} = 1.49$\ ($\Delta_{\mathrm{rms}}^{\mathrm{model}} \leq \Delta_{\mathrm{rms}}^{\mathrm{class}}$) and $P_{c/a}^{\mathrm{class}} = 1.34$\,per cent ($(c/a)^{\mathrm{model}} \leq (c/a)^{\mathrm{class}}$). Compared to normal distributions, these fractions are equivalent to a significance of $\approx 2.5\,\sigma$.

\subsection{Positions and orbital poles of the classical satellites}

Adding to this the requirement that the orbital poles of eight of the classical satellites are at least as strongly concentrated as observed reduces the frequencies substantially. Only $P^{\mathrm{class}}_{\mathrm{rms+poles}} = 0.012$\ (requiring $\Delta_{\mathrm{rms}}^{\mathrm{model}} \leq \Delta_{\mathrm{rms}}^{\mathrm{class}}$\ \textit{and} $\Delta_{\mathrm{std}}^{\mathrm{model}} \leq \Delta_{\mathrm{std}}^{\mathrm{class}}$) or $P^{\mathrm{class}}_{c/a\mathrm{+poles}} = 0.007$\,per cent (requiring $(c/a)^{\mathrm{model}} \leq (c/a)^{\mathrm{class}}$\ \textit{and} $\Delta_{\mathrm{std}}^{\mathrm{model}} \leq \Delta_{\mathrm{std}}^{\mathrm{class}}$) have as concentrated orbital poles and are simultaneously as flattened (measured using $\Delta_{\mathrm{rms}}$\ and $c/a$, respectively). This corresponds to a significance of 3.8 to $4.0\,\sigma$. We note that this is a conservative test because it ignores that the orbital poles are concentrated close to the normal direction to the VPOS (indicating the satellite structure is rotationally supported), and also that one of the remaining three classical satellites (Sculptor) is also orbiting within the VPOS, but in the opposite direction. Furthermore, the resulting frequencies do not change substantially (less than a factor of 2) if the number of considered most-concentrated orbital poles is changed from 8 to between 6 and 10.

\subsection{Positions of satellites discovered in SDSS}

How does the uneven SDSS footprint affect the flattening of the spatial distribution of SDSS satellites? Drawing 16 satellites (following the exact same radial distribution as the observed SDSS satellites) from an isotropic model for which only the region of $|b| < 24^{\circ}$\ is excluded (i.e. ignoring the SDSS footprint) results in an average rms height $<\Delta_{\mathrm{rms}}>$\ of $40.9$\,kpc and an average axis ratio of $<c/a> = 0.45$. Adding the requirement that the satellites are within the SDSS footprint (following the procedure outlined in Sect. \ref{fig:isofraction}) results in an average rms height of $<\Delta_{\mathrm{rms}}> = 33.3$\,kpc and an average axis ratio of $<c/a> = 0.37$. As expected, the survey footprint shape does indeed result in more narrow satellite distributions on average. But these average plane heights are still considerably larger than that of the observed 16 SDSS satellites of $\Delta_{\mathrm{rms}}^{\mathrm{SDSS}} = 25.9$\,kpc and $(c/a)^{\mathrm{SDSS}} = 0.26$.

In addition the typical orientation of the best-fit-planes (defined by their normal vector) is different from that of the observed SDSS satellite plane fit. Fig. \ref{fig:ASPnormals} shows the distribution the normal directions of these planes for 50000 realisations. The plot demonstrates that the uneven sky coverage of the SDSS footprint introduce a preferred orientation of plane normals. However, the preferred orientation for the SDSS-satellite plane normals (red contours) is different from that obtained for the plane fitted to the observed SDSS satellites (red diamond) by more than $30^{\circ}$. Furthermore, SDSS satellite planes at least as closely aligned with the classical satellite plane as observed (normal vectors aligned to $22^{\circ}$\ or better) are on average slightly wider ($<\Delta_{\mathrm{rms}}> = 34.7$\,kpc, $<c/a> = 0.40$) than those not as well aligned as observed ($<\Delta_{\mathrm{rms}}> = 33.1$\,kpc, $<c/a> = 0.37$). This indicates that the survey footprint, while biasing the satellite distribution towards a more narrow configuration, prefers an orientation which is different from the close alignment with the plane defined by the 11 classical satellites. This is also evidenced by the fraction of similarly closely aligned model satellite planes. If the model SDSS satellites are drawn from an isotropic distribution ignoring the SDSS footprint (but excluding $|b| < 24^{\circ}$) then 13.7\,per cent of the resulting planes are aligned at least as well as observed with the plane of the 11 classical satellites. However, this frequency drops to 7.6\,per cent when the SDSS satellites are required to be confined to the actual SDSS footprint (right panel of Fig. \ref{fig:results}). The survey shape thus clearly biases the SDSS satellites away from being similarly well aligned as observed instead of fostering an alignment as has often been argued.

The SDSS-equivalent model satellite distribution can be considered to be comparable to the observed situation if a plane fitted to their positions is at least as flattened (again measured either with the rms height or the axis ratio) and at least as closely aligned with the plane of classical satellites as observed. Of all 50000 isotropic realisations, $P^{\mathrm{SDSS}}_{\mathrm{rms}} = 0.82$\,per cent are at least as closely aligned with the classical satellite plane and have an as small $\Delta_{\mathrm{rms}}$\ as the observed value of $\Delta_{\mathrm{rms}}^{\mathrm{SDSS}} = 25.9$\,kpc. Replacing the $\Delta_{\mathrm{rms}}$\ criterion with the equivalent $c/a$\ criterion, thus requiring the axis ratio to be at least as extreme as observed, results in a frequency of $P^{\mathrm{SDSS}}_{c/a} = 0.53$\,per cent. If the distribution of the classical MW satellites is given, the flattening and alignment of only the SDSS satellite positions with the best-fit classical satellite plane is therefore already significant at the $> 99$\,per cent level (2.6 to $2.8\,\sigma$).

For the assumed isotropic model, the distribution of the classical satellites and of the SDSS satellites is statistically independent, so the overall significance of the VPOS consisting of both the classical and the SDSS satellites can be obtained by multiplying the frequencies to find the classical satellite distribution and the close alignment of the SDSS satellites with it. For the RMS height, $P^{\mathrm{VPOS}}_{\mathrm{rms}} = P^{\mathrm{class}}_{\mathrm{rms}} \times P^{\mathrm{SDSS}}_{\mathrm{rms}} = 9.8 \times 10^{-7}$\ (equivalent to $4.9 \sigma$), while for the axis ratio $P^{\mathrm{VPOS}}_{c/a} = P^{\mathrm{class}}_{c/a} \times P^{\mathrm{SDSS}}_{c/a} = 3.7 \times 10^{-7}$\ (equivalent to $5.1 \sigma$). Adding the information provided by the SDSS satellites thus increases the overall significance of the VPOS by about $1.1\,\sigma$. Contrary to commonly stated perceptions, the alignment of the faint satellites discovered in the SDSS survey with the VPOS thus does indeed add information on the VPOS phenomenon and is not merely an artefact of the SDSS footprint.

\subsection{Implications of a future all-sky survey}

\begin{figure}
   \centering
   \includegraphics[width=85mm]{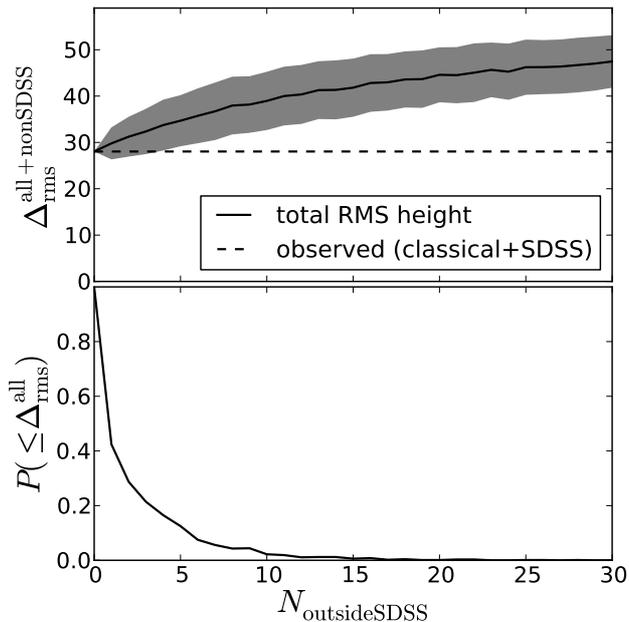}
   \caption{
\textit{Upper panel}: Average height (and standard deviation, grey area) of planes fitted to the 27 observed Milky Way satellite positions considered in this work, plus $N_\mathrm{outsideSDSS}$\ satellites drawn from isotropy but confined to the region outside of the SDSS survey and $b > 24^{\circ}$. 
\textit{Lower Panel}: The frequency that the model satellites lie within the VPOS, where ''within the VPOS'' is assumed to mean that the
fitted plane height including the model satellites is as narrow as the height of the plane fitted to only the observed positions of the 27 considered Milky Way satellites (indicated as the dashed line in the upper panel).
   }
              \label{fig:outsideSDSS}
\end{figure}

Motivated by the Pan-STARRS survey covering three quarters of the sky \citep{Kaiser2002}, and the so far surprisingly low number of objects discovered by it that can be unambiguously identified as MW satellite galaxies \citep{Laevens2013,Laevens2015a,Laevens2015b}, the following question arises: What if an all-sky survey with the same satellite detection limit as the SDSS would not detect a single additional satellite outside of the SDSS footprint? This would clearly increase the significance of the satellite anisotropy compared to an isotropic distribution, as it would imply that 16 out of 16 satellites lie within the SDSS footprint by chance, but none outside of it. Again assuming an isotropic satellite distribution, the chance to be within or outside of the SDSS footprint is given by the ratio of the area covered by the survey and the total available area. Assuming that the hypothetical all-sky survey has the same detection limits as the SDSS (and that the SDSS satellites are a representative sample) implies (1) that the radial satellite distribution is identical to that of that of the observed SDSS satellites, and (2) that the same Galactic latitude limit applies to the all-sky survey. Thus we assume that no satellite with $b \leq 24^{\circ}$\ can be discovered. This leaves $\sin(24^{\circ}) = 41$\,per cent of the sky free of model satellites. By generating 100000 random satellite positions drawn from isotropy and counting how many of these lie outside of the obscured region but either within or outside the SDSS, the relative areas of the SDSS footprint and its complement are determined. The counts show that the SDSS footprint occupies almost exactly half of the available sky after accounting for the adopted MW disk obscuration (30100 versus 29200 counts, which also implies the prediction that approximately 16 additional satellites should be discovered in an SDSS-like all-sky survey). Thus, the chance that all 16 out of 16 satellites drawn from isotropy end up within the SDSS area is approximately $0.5^{16} = 1.5 \times 10^{-5}$. Not detecting a single satellite outside of the SDSS region would thus substantially increase the significance of the satellite distribution.

This hypothetical situation already disagrees with our knowledge of the observed MW satellite system. Satellite galaxies and candidates have been discovered outside of the SDSS footprint \citep[e.g.][]{DES2015,Koposov2015,KimJerjen2015a,Martin2015,DES2015b}. However, most of those discoveries align with and thus confirm the previously identified VPOS \citep{Pawlowski2015a}. Thus, we may ask how the discovery of $N_\mathrm{outsideSDSS}$\ additional satellites outside of the SDSS footprint but within the VPOS affects its significance\footnote{If all new discoveries lie within the VPOS, the fraction $f_{\mathrm{outside}}$\ of satellites outside of the VPOS would of course decrease according to $f_{\mathrm{outside}} = N_{\mathrm{iso}}/(27+N_\mathrm{outsideSDSS})$.}. To determine the frequency with which the satellites align with the VPOS by chance, we proceed as follows. A number of $N_\mathrm{outsideSDSS}$\ satellites is drawn from the radial distance distribution of the SDSS satellites, which is equivalent to assuming that the survey has the same depth as the SDSS and that the radial distribution of the SDSS satellites is representative. Their positions on the sky are, as before, drawn from isotropy but are again required to lie at $b > 24^{\circ}$. However, this time the satellites are required to lie \textit{outside} of the SDSS footprint region. The upper panel of Figure \ref{fig:outsideSDSS} illustrates how this affects the expected thickness of a plane fitted to the 27 classical and SDSS satellites (at their observed positions) plus the additional $N_\mathrm{outsideSDSS}$\ model satellites. The additional satellites result in a pronounced widening of the satellite plane on average. Since the observed VPOS is not a clearly delineated structure, we adopt the following definition for ''all additional satellites lie in the VPOS'': they are distributed such that the overall satellite plane height does not increase. This allows to determine how the significance of the VPOS is affected if the additional satellites are found to lie within the observed VPOS. The lower panel of Figure \ref{fig:outsideSDSS} shows the fraction of realisations in which the resulting satellite plane height (measured using the 11 classical, 16 SDSS, and $N_\mathrm{outsideSDSS}$\ satellites in the compliment of the SDSS footprint but not obscured by the MW disk) is at least as small as the observed plane height for the current observed sample of the 11 classical and 16 SDSS satellites. The chance that the resulting satellite plane is as narrow as observed (i.e. the satellites are on average ''within the VPOS'') drops rapidly with $N_\mathrm{outsideSDSS}$. While it is above 10 per cent up until $N_\mathrm{outsideSDSS} = 5$, it is almost zero for $N_\mathrm{outsideSDSS}$\ beyond 10. Thus, an all-sky survey discovering 10 or more satellites outside of the SDSS but aligned with the VPOS would imply a substantial further increase of the significance of the structure.


\section{Fraction of isotropically distributed satellites}
\label{sect:isofraction}

\begin{figure*}
   \centering
   \includegraphics[width=170mm]{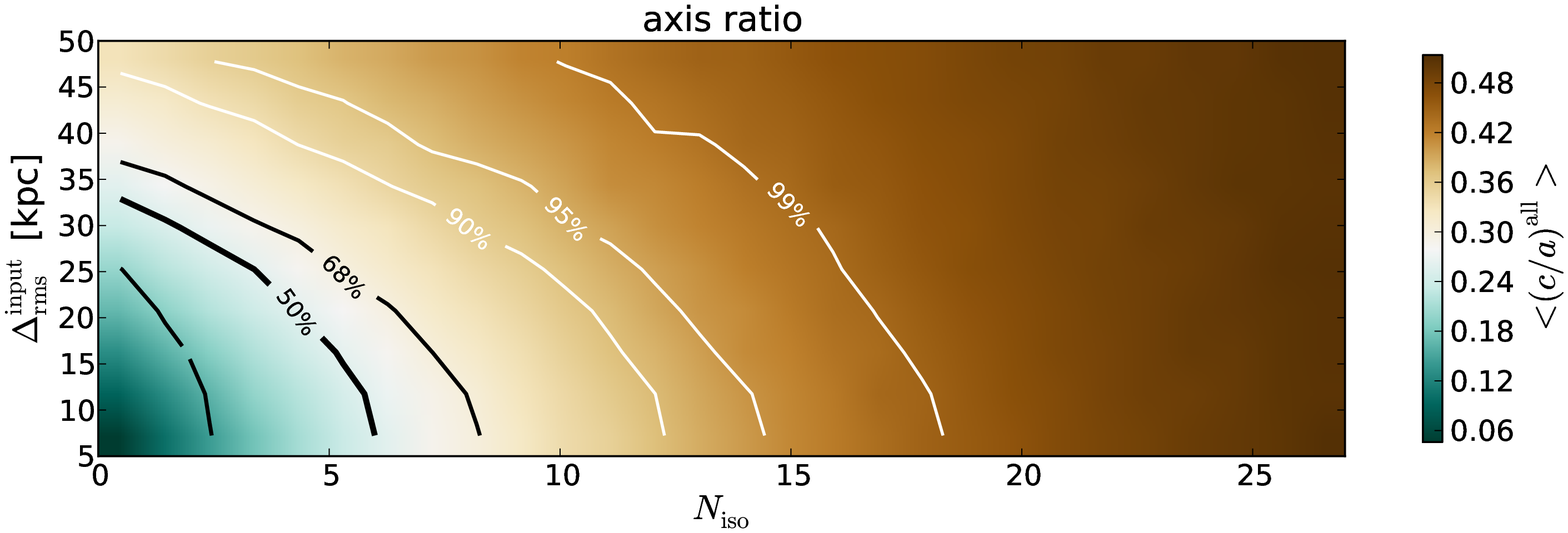}
   \includegraphics[width=170mm]{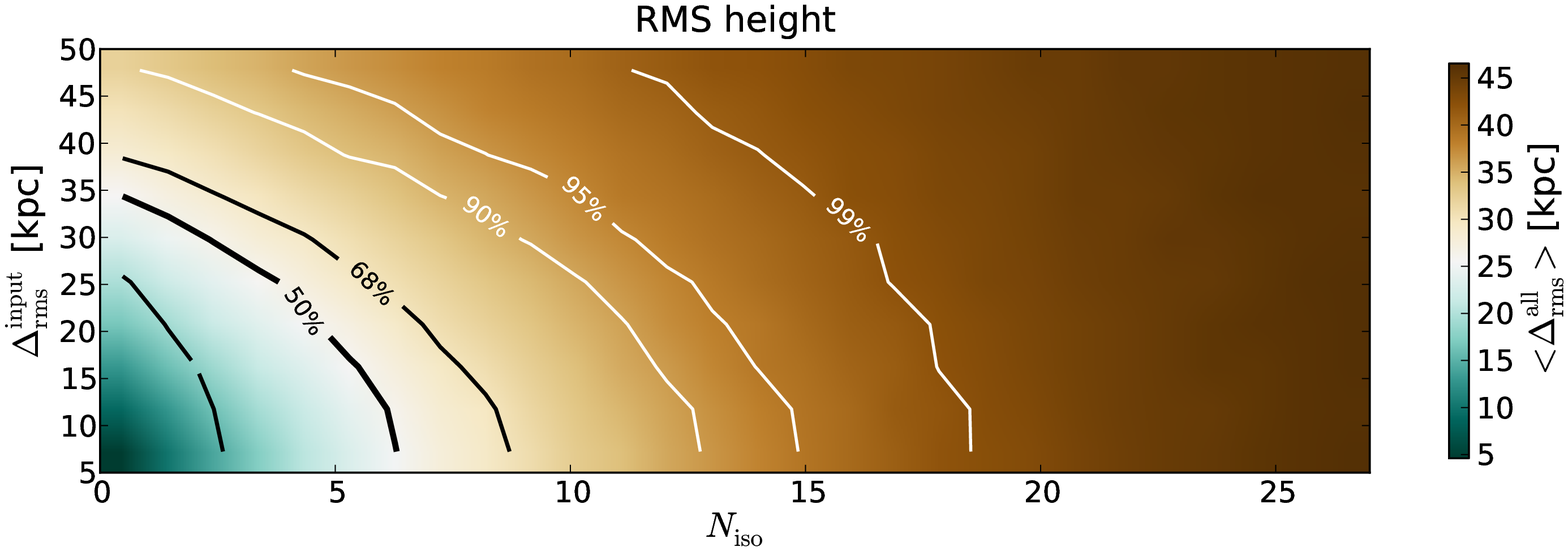}
   \caption{
Average axis ratio $c/a$\ (\textit{upper panel}) and rms height $\Delta_{\mathrm{rms}}$\ (\textit{lower panel}) for all combinations of input plane heights $\Delta_{\mathrm{rms}}^{\mathrm{input}}$\ and fraction of isotropically distributed satellites $N_{\mathrm{iso}}$, modelled with 5000 realisations each. The contours indicate where more than the specified fraction of realisations is less extreme than the VPOS. The expected number of isotropically distributed satellites (50\,per cent contour) ranges between 6 for very narrow input planes and 0 for input planes larger than 35\,kpc, and is about 2 for input plane heights comparable to the observed VPOS (between 25 and 30\,kpc). That more than about half of all 27 satellites are part of an isotropic distribution can be ruled out with high significance (95 to 99\,per cent).
   }
              \label{fig:isofraction}
\end{figure*}

The anisotropy of the MW satellite distribution was established beyond reasonable doubt in the previous section. We thus assume for the reminder of this contribution that at least some of the MW satellites are part of a planar distribution, and turn our attention to the question of which fraction of the considered satellites could nevertheless be part of an isotropic distribution. For simplicity and robustness we will ignore the information provided by the aligned orbital poles during this investigation, instead opting to use it later on as a consistency check of our results.

To estimate how many of the known MW satellites can belong to an isotropic distribution, we determine for which ratio of isotropically versus planar distributed satellites in model satellite systems we \textit{expect} an overall flattening as extreme as that of the observed MW satellite system. For this, artificial satellite distributions that are made up of $N_{\mathrm{iso}}$\ satellites drawn from an isotropic distribution, and of 27-$N_{\mathrm{iso}}$\ satellites drawn from a planar distribution with an input plane height of $\Delta_{\mathrm{rms}}^{\mathrm{input}}$\ are constructed as described in Sect. \ref{sect:modelplanes}. 

Figure \ref{fig:isofraction} summarizes the results. The average axis ratio $<(c/a)^{\mathrm{all}}>$\ or rms height $<\Delta_{\mathrm{rms}}^{\mathrm{all}}>$\ over all 5000 realisations for each combination of input plane width $\Delta_{\mathrm{rms}}^{\mathrm{input}} = [5, 10, ..., 45, 50]$\ and number of isotropically distributed satellites $N_{\mathrm{iso}} = [1,2,3, ... ,25, 26, 27]$\ are colour coded. The contours indicate the fraction of systems which are less extreme than the observed MW satellite system (e.g. larger $(c/a)^{\mathrm{all}}$\ or larger $\Delta_{\mathrm{rms}}^{\mathrm{all}}$). The results are essentially identical for the axis ratio and rms height plots.

Not surprisingly, a larger $N_{\mathrm{iso}}$\ results in less narrow planes. The typical axis ratio for a fully isotropic MW satellite system (considering the anisotropy introduced by the MW obscuration for the 11 classical satellites and the uneven survey footprint for the SDSS satellites) is $<(c/a)^{\mathrm{all}}> (N_{\mathrm{iso}}) \approx 0.5$, and the typically expected rms height is $\Delta_{\mathrm{rms}}^{\mathrm{all}}(N_{\mathrm{iso}}) \approx 46\,\mathrm{kpc}$. Both values are almost twice as large as those determined from the observed satellite distribution.

For lower $N_{\mathrm{iso}}$, the contribution of on-plane satellites becomes increasingly important. For $N_{\mathrm{iso}} \rightarrow 0$, $<\Delta_{\mathrm{rms}}^{\mathrm{all}}>$\ approaches $\Delta_{\mathrm{rms}}^{\mathrm{input}}$\ for small input plane heights. For larger input heights the measured heights tend to be smaller than the input heights, which can be attributed to the non-complete sky coverage and the requirement that a satellite's offset from the input plane can not be larger than its Heliocentric distance (which is required to keep the Galactocentric radial distribution of model satellites identical to the observed one).

The fraction of observed MW satellites which are part of an isotropic populations can be estimated by finding that $N_{\mathrm{iso}}$\ where the chance to have a similarly extreme satellite plane as observed for the MW exceeds 50\,per cent -- or alternatively where the average plane height is equal to that of the observed VPOS. As can be seen in Fig. \ref{fig:isofraction} both methods effectively result in the same answer. Depending on the input plane height, we expect that two ($\Delta_{\mathrm{rms}}^{\mathrm{input}} = 30\,\mathrm{kpc}$) to six ($\Delta_{\mathrm{rms}}^{\mathrm{input}} = 10\,\mathrm{kpc}$) of the 27 considered MW satellite galaxies are part of an isotropic distribution. That more than 50\,per cent (14 of 27) of these satellites are part of an isotropic contribution can be excluded at 95 to 99\,per cent confidence.

These results are well comparable to the GPoA which consists of about half of all M31 satellites. According to our results, an isotropic contribution of more than $\sim 50$\,per cent is essentially excluded, the VPOS is instead expected to consist of $\geq 80$\,per cent of the MW satellites considered in this study. This fraction is compatible with the fraction of MW satellites which \citet{PawlowskiKroupa2013} found to be consistent with orbiting within the VPOS (9 of 11, with 8 of these co-orbiting), and also with the number of outliers (three out of 27) from the best-fit VPOS plane discussed in \citet{Pawlowski2013}. Taken together, these two studies suggest that the following four MW satellites are likely not part of the VPOS: Sagittarius (not orbiting in the VPOS), Leo I (large offset from the best-fit plane and possibly not orbiting in the VPOS\footnote{We caution against overrating the orbital pole of Leo I. There is only one PM measurement for Leo I yet \citep{Sohn2013}, and for other objects different measurements sometimes differ by much more than the reported uncertainties, indicating that the proper motion errors are dominated by underlying systematics. Furthermore, Leo I has a very large radial velocity relative to the Galactic centre and might not even be a satellite dynamically bound to the Galaxy if the latter is of low mass \citep{Sohn2013,BoylanKolchin2013}.}), Ursa Major I and Hercules (both have a large offset from the best-fit plane).
In case of M31, the number of known off-plane satellites might be larger than for the MW because large parts of the region obscured by the MW disk are outside of the VPOS plane, whereas for M31 the region close to the galaxy's disk has also been covered by PAndAS (and in fact a number of M31 satellites lie close to the plane defined by M31's stellar disk, see \citealt{Pawlowski2013}). However, because our method explicitly excludes the region obscured by the MW, the potential discovery of such objects does not affect our determined significance.


\section{Discussion and Conclusion}
\label{sect:conclusion}

For the first time, we have calculated the significance of the VPOS including the fainter MW satellites discovered in the SDSS. For this purpose, we have characterized the VPOS by the spatial flattening and orbital coherence of the 11 classical satellite and the similar spatial alignment of fainter satellite galaxies discovered in the SDSS survey. We determined the frequency to find similar coherences in isotropic distributions that follow the exact same (Galactocentric) radial distribution as the observed MW satellite galaxies. 

That the alignment of the 16 considered SDSS satellites in a narrow plane -- which is oriented similar to the plane fitted to the classical satellite galaxies -- is due to the survey footprint shape can be ruled out at more than 99\,per cent confidence. We find that the phase-space coherence of the 11 classical satellites alone already results in a very high significance of the VPOS ($\approx 3.9\,\sigma$), constituting good evidence for the existence of the structure. Adding the information provided by the fainter satellites discovered in the SDSS by modelling the exact survey footprint and taking a stronger obscuration by the MW disk for these fainter objects into account, we find that the significance of the VPOS increases to $\approx 5\,\sigma$. Not only does this essentially rule out the possibility that the VPOS is a mere chance alignment, it actually constitutes a discovery of the VPOS by the stringent standards of high-energy physics.

How does the significance of the VPOS compare with that of the similar structure around M31, the Great Plane of Andromeda (GPoA) discovered by \citet{Ibata2013}? Comparing with distributions drawn from isotropy but confined by the survey constraints, they report a significance of 99.87\,per cent (equivalent to $3.2 \sigma$) for the spatial alignment alone, and of 99.998\,per cent (equivalent to $\sim 3.7 \sigma$) if the line-of-sight velocity coherence is also taken into account. This means that the VPOS is more significant than the GPoA. Both analyses are based on the same number of satellites (27), but in the MW case the existence of proper motion measurements for the 11 classical satellites provides important additional information on the phase-space alignment of the objects. In fact, the significance of the 11 classical satellites alone already surpasses that of the GPoA.

Motivated by the fact that the GPoA consists of only half of the M31 satellite galaxies, we have also determined which fraction of the 27 considered MW satellite galaxies is expected to belong to an isotropic distribution. Depending on the underlying rms height $\Delta_{\mathrm{rms}}^{\mathrm{input}}$\ of the VPOS, we expect between two ($\Delta_{\mathrm{rms}}^{\mathrm{input}} = 30$\,kpc) and six ($\Delta_{\mathrm{rms}}^{\mathrm{input}} = 10$\,kpc) of the considered satellites to be part of an isotropic distribution. A contribution of more than about 50\,per cent of isotropically distributed satellites can be ruled out at 95 to 99\,per cent confidence.

We can thus conclude that, under the assumption of an isotropic null-hypothesis, the current distribution of MW satellites is so unexpected to arise by chance that this possibility constitutes, with \citet{Bernoulli1734}'s words, a ''impossibilit\'e morale''\footnote{"moral impossibility"}. The high significance of the VPOS thus does reveal that its existence needs an explanation -- in any cosmological framework.
No comparison to satellite distributions expected in the $\Lambda$CDM cosmology were performed in this work, so the presented results do not immediately affect that theory. Populations of $\Lambda$CDM sub-halos are known to be anisotropic due to correlations in the infall of structures onto host halos, such as the preferential accretion along dark matter filaments \citep{Zentner2005, Libeskind2015} and the infall of sub-halos in groups \citep{DOnghiaLake2008,Wetzel2015}. However, overall the flattening of sub-halo populations shows pronounced similarities to isotropic distributions \citep{Pawlowski2012b, PawlowskiKroupa2013, PawlowskiMcGaugh2014b}. Previous studies comparing the observed and expected phase-space distribution of satellite population mostly focussed on the 11 brightest MW satellite galaxies, and already found a severe mismatch with dark matter only simulations\citep{Pawlowski2014,PawlowskiMcGaugh2014b}. Neither has the inclusion of baryonic effects in cosmological simulations yet been shown to reduce the mismatch \citep{Pawlowski2015b}. In a future publication we plan to use the technique of applying the exact SDSS coverage developed for this paper to modelled satellite distributions in $\Lambda$CDM simulations. This will reveal whether the SDSS satellites compound the discrepancy between the observed MW satellite galaxies and expectations derived from $\Lambda$CDM simulations.


\section*{Acknowledgements}
MSP thanks Stacy McGaugh, Federico Lelli and Heather Morrison for helpful comments, Vasily Belokurov for a detailed discussion during the 11th Potsdam Thinkshop which raised the question addressed in Sect. \ref{sect:isofraction}, and David Merritt for pointing out the work of Daniel Bernoulli. This publication was made possible through the support of a grant from the John Templeton Foundation.

\bibliographystyle{mnras}
\bibliography{isobib}

\bsp	
\label{lastpage}
\end{document}